# Rapid-convergent nonlinear differentiator

## Xinhua Wang, Bijan Shirinzadeh

Robotics and Mechatronics Research Laboratory, Department of Mechanical and Aerospace Engineering,

Monash University, Clayton, VIC, 3800, Australia

E-mail: wangxinhua04@gmail.com

**Abstract:** A nonlinear differentiator being fit for rapid convergence is presented, which is based on singular perturbation technique. The differentiator design can not only sufficiently reduce the chattering phenomenon of derivative estimation by introducing a continuous power function, but the dynamical performances are also improved by adding linear correction terms to the nonlinear ones. Moreover, strong robustness ability is obtained by integrating nonlinear items and the linear filter. The merits of the rapid-convergent differentiator include the excellent dynamical performances, restraining noises sufficiently, avoiding the chattering phenomenon and being not based on system model. The theoretical results are confirmed by computer simulations and an experiment.

## 1. Introduction

Differentiation of signals is an old and well-known problem [1, 2, 3] and has attracted more attention [4, 5, 6] in recent years. Rapidly and accurately obtaining the velocities and accelerations of tracked targets is crucial for several kinds of systems with correct and timely performances, such as the missile-interception systems in defence systems [22] and underwater vehicle systems [23]. Therefore, the convergence rates, robustness and computation time of differentiators are very significant.

Difference method is used to estimate approximately the derivatives of signals. Because disturbances exist in almost all the signals, the satisfying estimation precision cannot be obtained by difference method. Kalman filter can reduce the disturbance and obtain the derivatives of signals. However, the model of plant must be required. This constrains the types of signals. Therefore, it is very important to construct an estimator with filtering ability and without knowing the model of plant.

Usually the uncertainties of nonlinear systems are approximated by some intelligent algorithms, such as fuzzy systems or radial-based-function (RBF) networks. Over the last decade, control researchers tried to apply intelligent methodologies to the design of estimation and control for uncertain nonlinear systems [7]-[17]. In paper [7], a conclusion was given that fuzzy systems are universal approximation. In paper [7], the Stone-Weierstrass Theorem is used to prove that fuzzy systems with product inference, centroid defuzzification, and Gaussian membership function are capable of approximating any real continuous function on a compact set to arbitrary accuracy. Paper [8] reported on a related study of radial-basis-function (RBF) networks, and it was proved that RBF networks having one hidden layer are capable of universal approximation. The main difficulties with estimating uncertainties by these algorithms are: 1) the parameters or the neural network weights are difficult to be regulated; 2) membership function and Gaussian function are selected by experiences; 3) all of the system states must be required for estimation and feedback; 4) noises cannot be restrained by the above approximation algorithms. In [18, 19, 20], the differential transformation method (DTM) and the finite difference method (FDM) are used to analyze the nonlinear dynamic behavior of a rigid rotor supported by a micro gas bearing system (MGBS), and in [21], a nonlinear rule based fuzzy logic controller is proposed to control the nonlinear dynamic behavior of the probe tip of an atomic force microscope system (AFMs). However, the abilities of restraining noise were not considered.

Differentiators are usually used to obtain the information of tracked targets or the information of the plants which are not modeled easily. For instance, the position of a tracked target is measured by radar, and its velocity can be carried out by a differentiator. Differentiators are the estimators which are not based on the model of plant. The popular high-gain differentiators [4] provide for an exact derivative when their gains tend to infinity. Unfortunately, their sensitivity to small high-frequency noise also infinitely grows. With any finite gain values such a differentiator has also a finite bandwidth. Thus, being not exact, it is, at the same time, insensitive with respect to high-frequency noise. Such insensitivity may be considered both as advantage or disadvantage depending on the circumstances. In [5, 6], a differentiator via second-order (or high-order) sliding modes algorithm has been proposed. The information one needs to know on the signal is an upper bound for Lipschitz constant of the derivative of the signal. Although second-order sliding mode is introduced and there exists no chattering phenomenon in signal tracking. However, for the robust exact differentiator, in the second dynamical equation, a switching function exists. The output of derivative estimation is continuous but not smooth. Therefore, chattering phenomenon still exists in derivative estimation. In [34], we presented a finite-time-convergent differentiator that is based on singular perturbation technique [24-29]. The merits of this differentiator exist in three aspects: rapidly finite-time convergence compared with other typical differentiators; no chattering phenomenon; and besides the derivatives of the derivable signals, the generalized derivatives of some classes of signals can be obtained. However, the differentiators in [34] are complicated and difficult to be used in practice. Moreover, the computation time is still large and its robustness is not considered.

In this paper, we expand our work in [34]. The proposed rapid-convergent differentiator is an integration of a nonlinear term (comprising of continuous power function) and a linear correction term. The structure of the designed differentiator is simple and the computation time is short. In the nonlinear part, a continuous power function with a perturbation parameter is used. Therefore, chattering phenomenon can be avoided in the output of derivative estimation. Strong robustness ability is obtained by integrating sliding mode items and the linear filter. The linear part can restrain high-frequency noises by giving a suitable nature frequency, and small bounded noises can be restrained by the nonlinear items. Moreover, the dynamical performances are also improved.

This paper is organized in the following format. In section 2, problem statement is given, and preliminaries are given in section 3. In section 4, our main results are presented. In section 5, the main results are proved in detail. In section 6, robustness analysis of finite-time-convergent differentiator is given. In section 7, reduction of peaking phenomenon is given. In section 8, the simulations and experiments are given, and our conclusions are made in section 9.

## 2. Problem statement

When we design a controller for a system, state feedback is used to place the poles. In order to use the system states to design feedback controller, sensors must be adopted to measure the states. However, not all states can be measured directly. Therefore, state observer is presented to estimate the system states. These observers are designed under the condition that the system model must be known.

We present the following three questions.

**Question 1:** For arbitrary signals which cannot be expressed by mathematical expressions, such as the signal shown in Fig. 1, how can we estimate the derivate of the signal?

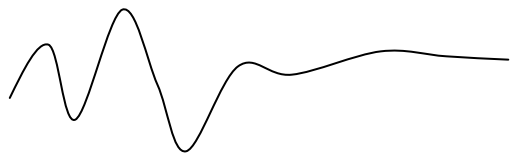

Fig. 1 Stochastic signal

**Question 2:** How can we obtain the derivative of a signal with noise? For instance, the following signal in Fig. 2.

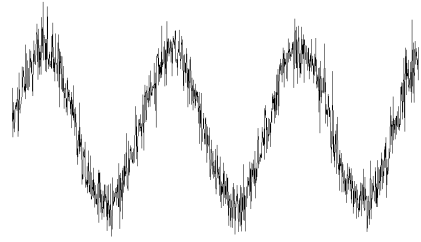

Fig. 2 Signal with noises

**Question 3:** For system

$$\dot{x}_1 = x_2$$
$$\dot{x}_2 = u \qquad (1)$$
$$y = x_1$$

we are required that the output $y$ tracks the desired trajectory $y_d(t)$ shown in Fig. 3.

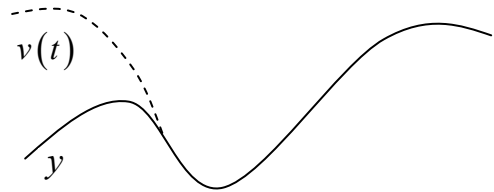

Fig. 3 Trajectory tracking

Let

$$e_1 = x_1 - y_d(t), e_2 = x_2 - \dot{y}_d(t) \qquad (2)$$

we get the tracking error system as follow:

$$\dot{e}_1 = \dot{e}_2$$
$$\dot{e}_2 = u - \ddot{y}_d \qquad (3)$$

Before designing contoller $u$, how can we obtain the information of $\dot{y}_d$ and $\ddot{y}_d$?

Design of usual observer is based on the model of system. However, most signals are difficult to be described by a given model. Therefore, using usual observer to estimate derivative of signal is restricted.

We expect to construct the following form:

$$\dot{x}_1 = x_2$$
$$\dot{x}_2 = f(x_1 - v(t), x_2) \qquad (4)$$

Under the condition that the solution exists for the above differential equation, a differentiator can be constructed if the following relation is satisfied: $x_1$ converges to $v(t)$ and $x_2$ converges to $\dot{v}(t)$.

We are interested in designing a rapid-convergent differentiator to estimate the derivative of signal, and high-frequency noise can be restrained sufficiently.

## 3. Preliminaries

First of all, the related concepts are given. In order to demonstrate the finite-time convergence, we introduce the following definitions, assumptions and theorems from the corresponding references.

**Definition 1 [35].** "Generalized derivative" denotes that the left and right derivatives of a point in a function trajectory both exist, and they may be not equal to each other.

**Definition 2 [33].** Consider a time-invariant system in the form of

$$\dot{x} = f(x), f(0) = 0, x \in \Re^n \tag{5}$$

where $f : D \to \Re^n$ is continuous on open neighborhood $D \subset \Re^n$ of the origin. The origin is said to be a finite-time-stable equilibrium of the above system if there exists an open neighborhood $N \subseteq D$ of the origin and a function $T_f : N \setminus \{0\} \to (0, \infty)$, called the settling-time function, such that the following statements hold:

1) *Finite-time-convergence:* For every $x \in N \setminus \{0\}$, $\psi^x$ is the flow starting from $x$ and defined on $[0, T_f(x))$, $\psi^x(t) \in N \setminus \{0\}$ for all $t \in [0, T_f(x))$, and $\lim_{t \to T_f(x)} \psi^x(t) = 0$.

2) *Lyapunov stability:* For every open neighborhood $U_\varepsilon$ of zero, there exists an open subset $U_\delta$ of $N$ containing zero such that, for every $x \in U_\delta \setminus \{0\}$, $\psi^x(t) \in U_\varepsilon$ for all $t \in [0, T_f(x))$.

The origin is said to be a globally finite-time-stable equilibrium if it is a finite-time-stable equilibrium with $D = N = \Re^n$. Then the system is said to be finite-time-convergent with respect to the origin.

**Assumption 1.** Suppose the origin is a finite-time-stable equilibrium of [33, Theorem 4.3] of

$$\begin{aligned}
\dot{\tilde{z}}_1 &= \tilde{z}_2 \\
&\cdots \\
\dot{\tilde{z}}_{n-1} &= \tilde{z}_n \\
\dot{\tilde{z}}_n &= f(\tilde{z}_1, \tilde{z}_2, \cdots, \tilde{z}_n)
\end{aligned} \tag{6}$$

And the settling-time function $T_f$ is continuous at zero, where $f(\cdot)$ is continuous and $f(0)$. Let $N$ be as in Definition 2 and let $\theta \in (0,1)$. Then there exists a continuous scalar function $V$ such that 1) $V$ is positive definite and 2) $\dot{V}$ is real valued and continuous on $N$ and there exists $c > 0$ such that

$$\dot{V} + cV^\theta \leq 0 \qquad \qquad \text{for any } \theta \in (0,1) \tag{7}$$

**Assumption 2.** There exists a Lipschitz Lyapunov function $V$ satisfying (4) with Lipschitz constant $M$.

**Assumption 3.** For (6), there exist $\rho_i \in (0,1], i = 0, 1, \cdots, n-1,$ and a nonnegative constant $\bar{a}$ such that

$$\left| f(\tilde{z}_1, \tilde{z}_2, \cdots, \tilde{z}_n) - f(\bar{z}_1, \bar{z}_2, \cdots, \bar{z}_n) \right| \leq \bar{a} \sum_{i=1}^{n} |\tilde{z}_i - \bar{z}_i|^{\rho_i - 1} \tag{8}$$

where $\tilde{z}_i, \bar{z}_i \in \Re, i = 1, \cdots, n$.

**Assumption 4.** $v(t)$ is a continuous and piecewise $n$-order derivable signal with the following properties. The derivatives of $v(t)$ up to order $n$-2 exist on the whole time domain and $v(t)$ is not ($n$-1)-order differentiable at some instants $t_j$, $j=1,\cdots,k$. However, the $(n-1)$-order left derivative $v_-^{(n-1)}(t_j)$ and right derivative $v_+^{(n-1)}(t_j)$ exist [35], respectively, and $v_-^{(n-1)}(t_j) \neq v_+^{(n-1)}(t_j)$, $j=1,\cdots,k$, may be satisfied.

**Remark 1.** There are a number of nonlinear functions actually satisfying this assumption. For example, one such function is $x^{\rho_i}$ since $\left| x^{\rho_i} - \bar{x}^{\rho_i} \right| \leq 2^{1-\rho_i} |x-\bar{x}|^{\rho_i}, \rho_i \in (0,1]$. Moreover, there are smooth functions also satisfying this property. In fact, it is easy to verify that $\left| \sin x - \sin \bar{x} \right| \leq 2|x-\bar{x}|^{\rho_i}$ for any $\rho_i \in (0,1]$.

**Theorem 3.1 in [30].** Let the differential equation

$$\dot{z}(\tau) = f(z_1(\tau),\cdots,z_n(\tau)) \tag{9}$$

be locally homogeneous of degree $q < 0$ with respect to dilation $(r_1, \ldots, r_n)$ and let the equilibrium point $z = 0$ of (9) be globally uniformly asymptotically stable. Then the differential equation (9) is globally uniformly finite time stable.

**Theorem 1 in [34].** If

$$
\begin{aligned}
\dot{\tilde{z}}_1 &= \tilde{z}_2 \\
&\cdots \\
\dot{\tilde{z}}_{n-1} &= \tilde{z}_n \\
\dot{\tilde{z}}_n &= f(\tilde{z}_1, \tilde{z}_2, \cdots, \tilde{z}_n)
\end{aligned}
\tag{10}
$$

is satisfied with Assumptions 1-3, signal $v(t)$ is satisfied with Assumption 4. Then for system

$$
\begin{aligned}
\frac{dx_1}{dt} &= x_2 \\
&\cdots \\
\frac{dx_{n-1}}{dt} &= x_n \\
\varepsilon^n \frac{dx_n}{dt} &= f(x_1 - v(t), \varepsilon x_2, \cdots, \varepsilon^{n-1} x_n)
\end{aligned}
\tag{11}
$$

there exist $\gamma > 0$ (where $\rho\gamma > n$) and $\Gamma > 0$ such that

$$x_i - v^{(i-1)}(t) = O\left(\varepsilon^{\rho\gamma-i+1}\right) \tag{12}$$

for $t_j > t \geq t_{j-1} + \varepsilon\Gamma(\Xi(\varepsilon)e_+(t_{j-1}))$, $j = 1,\cdots,k+1$, (let $t_0 = 0$ and $t_{k+1} = \infty$ ) $i = 1,\cdots,n$ with

$$x_n(t_j) - v_-^{(n-1)}(t) = O\left(\varepsilon^{\rho\gamma-n+1}\right), j = 1,\cdots,k \tag{13}$$

where $\varepsilon > 0$ is the perturbation parameter and $O\left(\varepsilon^{\rho\gamma-i+1}\right)$ denotes the approximation of $\varepsilon^{\rho\gamma-i+1}$ order [28] between

$x_i$ and $v^{(i-1)}(t)$; and $\gamma = (1-\theta)/\theta$, $\theta \in (0,\min\{\rho/(\rho+n),1/2\})$, $n \geq 2$. $e_i = x_i - v^{(i-1)}, i=1,\cdots,n$. $e = [e_1 \quad \cdots \quad e_n]^{\mathrm{T}}$, $e_+(t_{j-1}) = [e_1(t_{j-1}) \quad \cdots \quad e_{n-1}(t_{j-1}) \quad e_n^+(t_{j-1})]^{\mathrm{T}}$, $e_n^+(t_{j-1}) = x_n - v_+^{(n-1)}(t_{j-1})$, and $\Xi(\varepsilon) = diag\{1,\varepsilon,\cdots,\varepsilon^{n-1}\}$.

The practical form of differentiator in [34] is:

$$\dot{x}_1(t) = x_2(t)$$
$$\varepsilon^2 \dot{x}_2(t) = u$$
$$u = -sat_{\varepsilon_b}\left\{sign\left(\phi_\alpha(x_1 - v(t), \varepsilon x_2)\right)\left|\phi_\alpha(x_1 - v(t), \varepsilon x_2)\right|^{\alpha/(2-\alpha)}\right\} - sat_{\varepsilon_b}\left\{sign(x_2)\left|\varepsilon x_2\right|^\alpha\right\}$$

where

$$\phi_\alpha(x_1 - v(t), \varepsilon x_2) = x_1 - v(t) + \frac{sign(x_2)\left|\varepsilon x_2\right|^{2-\alpha}}{2-\alpha}, \quad sat_{\varepsilon_b}(x) = \begin{cases} x, & |x| < \varepsilon_b \\ \varepsilon_b sign(x), & |x| \geq \varepsilon_b \end{cases}$$

Its structure is too complicated. In the following, we will design more simple differentiators based on Theorem 1 in [34], and the designed rapid-convergent differentiator can keep more rapidly convergent at all times.

**Denotations:**

$sig(y)^\alpha = |y|^\alpha \, \text{sgn}(y), \quad \alpha > 0$. It is obvious that $sig(y)^\alpha = y^\alpha$ only if $\alpha = q/p$ where $p, q$ are positve odd numbers. Moreover,

$$\frac{d}{dy}|y|^{\alpha+1} = (\alpha+1)sig(y)^\alpha, \quad \frac{d}{dy}sig(y)^{\alpha+1} = (\alpha+1)|y|^\alpha \tag{14}$$

In the following, we will give our main results of rapid-convergent differentiator.

## 4. Rapid-convergent differentiator

Here, we will design a nonlinear differentiator being fit for rapid convergence. The proposed differentiator consists of linear and nonlinear parts. It is shown that for arbitrary signal, there exists a differentiator achieving finite-time error decay. Moreover, high-frequency noise can be restrained sufficiently.

We design a rapid-convergent differentiator as follow:

**Theorem 1.** The rapid-convergent differentiator is designed as:

$$\dot{x}_1 = x_2$$
$$\varepsilon^2 \dot{x}_2 = -a_{10}(x_1 - v(t)) - a_{11}sig(x_1 - v(t))^{\frac{\alpha}{2-\alpha}} - a_{20}\varepsilon x_2 - a_{21}sig(\varepsilon x_2)^\alpha \tag{15}$$
$$y = x_2$$

For a continuous and piecewise two-order derivative signal $v(t)$, there exists $\gamma > 0$ (where $\rho\gamma > 2$ and $\rho = \min\{\alpha, \alpha/(2-\alpha)\} = \alpha/(2-\alpha)$), such that

$$x_i - v^{(i-1)}(t) = O\left(\varepsilon^{\rho\gamma-i+1}\right) \tag{16}$$

for $t \geq \varepsilon\Gamma\left(\Xi(\varepsilon)e(0)\right), i = 1$, and $t_j > t \geq t_{j-1} + \varepsilon\Gamma\left(\Xi(\varepsilon)e_+(t_{j-1})\right), j = 1, \cdots, k+1$, $i = 2$, respectively, with $x_2(t_j) - v_-'(t_j) = O\left(\varepsilon^{\rho\gamma-1}\right), j = 1, \cdots, k$, where $\alpha \in (0,1)$, $\varepsilon > 0$ is the perturbation parameter and $O\left(\varepsilon^{\rho\gamma-i+1}\right)$ denotes the approximation of $\varepsilon^{\rho\gamma-i+1}$ order [28] between $x_i$ and $v^{(i-1)}(t)$; $e_i = x_i - v^{(i-1)}, i = 1,2$. $e = \begin{bmatrix} e_1 & e_2 \end{bmatrix}^T$, $e_+(t_{j-1}) = \begin{bmatrix} e_1(t_{j-1}) & e_2^+(t_{j-1}) \end{bmatrix}^T$, $e_2^+(t_{j-1}) = x_2 - v_+'(t_{j-1})$, and $\Xi(\varepsilon) = diag\{1, \varepsilon\}$.

Equation (16) denotes that variable $x_i$ track ($i$-1)-order derivative of signal $v(t)$.

The rapid-convergent differentiator (15) consists of linear and nonlinear differentiator given, respectively, as follows

$$\dot{x}_1 = x_2$$
$$\varepsilon^2 \dot{x}_2 = -a_{10}\left(x_1 - v(t)\right) - a_{20}\varepsilon x_2 \qquad (17)$$
$$y = x_2$$

and

$$\dot{x}_1 = x_2$$
$$\varepsilon^2 \dot{x}_2 = -a_{11}sig\left(x_1 - v(t)\right)^{\frac{\alpha}{2-\alpha}} - a_{21}sig\left(\varepsilon x_2\right)^{\alpha} \qquad (18)$$
$$y = x_2$$

For linear differentiator (17), we have the following conclusion:

**Theorem 2:** For linear differentiator (17) and a signal $v(t)$ (where $v(t)$ is a continuous and piecewise second-order derivable signal, we have that

$$x_2(t) - \dot{v}(t) = O(\varepsilon) \qquad \text{for} \quad t \to \infty \qquad (19)$$

where $a_{10}$, $a_{20} > 0$, $\varepsilon > 0$ is perturbation parameter, $O(\varepsilon)$ denotes the approximation of $\varepsilon$ order [28] between $x_2$ and $\dot{v}(t)$.

For nonlinear differentiator (18), we have the following conclusion:

**Theorem 3:** For nonlinear differentiator (18) and a continuous and piecewise two-order derivative signal $v(t)$, there exists $\gamma > 0$ (where $\rho\gamma > 2$ and $\rho = \min\{\alpha, \alpha/(2-\alpha)\} = \alpha/(2-\alpha)$,), such that

$$x_i - v^{(i-1)}(t) = O\left(\varepsilon^{\rho\gamma-i+1}\right) \qquad (20)$$

for $t \geq \varepsilon\Gamma\left(\Xi(\varepsilon)e(0)\right), i=1$ , and $t_j > t \geq t_{j-1} + \varepsilon\Gamma\left(\Xi(\varepsilon)e_+\left(t_{j-1}\right)\right), j=1,\cdots,k+1$ , $i=2$ , respectively, with $x_2\left(t_j\right) - v'_-\left(t_j\right) = O\left(\varepsilon^{\rho\gamma-1}\right), j=1,\cdots,k$ , where $\alpha \in (0,1)$.

**Remark 2:** From the analysis of linear systems and non-Lipschitz nonlinear systems, linear differentiator (17) and nonlinear differentiator (18), two notable differences can be observed. First, the two algorithms have quite different converging properties: the linear system converges exponentially, whereas the trajectories of the non-Lipschitz nonlinear algorithm converge in finite time. This is due to the lack of local Lipschitzness of the nonlinear algorithm at the origin, that is, its behavior around the zero state is very strong compared to the linear case. On the other side, the linear correction terms are stronger than the ones of the nonlinear algorithm far from the origin. These differences cause another striking difference between both algorithms: the kind of perturbations that each one is able to tolerate. The main difference is that the linear system can deal with perturbations that are stronger very far from the origin and weaker near the origin than the ones that are endured by the nonlinear algorithm. So, for example, the nonlinear algorithm is not able to endure (globally) a bounded perturbation with linear growth in time, but the linear algorithm can deal with it easily. However, the linear algorithm is not able to support a strong perturbation near the origin, what is one of the main advantages of non-Lipschitz nonlinear algorithm.

In order to integrate the merits of linear differentiator (17) and nonlinear differentiator (18), and to restrain their shortcomings respectively, rapid-convergent differentiator (15) is integrated. The proposed rapid-convergent differentiator is an integration of a nonlinear term (comprising of continuous power function) and a linear correction term. The overall design is an integration of nonlinear non-Lipschitz differentiators (18) and linear differentiator (17) in that a perturbation parameter α is introduced which takes values (0,1) and an additional linear correction term is appended, so that it inherits the best properties of both. In the nonlinear part, a continuous power function with a perturbation parameter is used. Therefore, chattering phenomenon can be avoided in the output of derivative estimation. Strong robustness ability is obtained by integrating sliding mode items and the linear filter. The linear part can restrain high-frequency noises by giving a suitable nature frequency, and small bounded noises can be restrained by the nonlinear items, at the same time, the nonlinear items can compensate the delay brought by the linear filter.

## 5. Proof of Theorems 1-3.

### 5.1 Rapid convergence

In order to analyse differentiators (15), (17) and (18), firstly, we give the following system convergence lemmas 1-3.

**Lemma 1** (Convergence of linear system)**.** System

$$\begin{cases} \dfrac{d(z_1(\tau))}{d\tau} = z_2(\tau) \\ \dfrac{d(z_2(\tau))}{d\tau} = -a_{10}z_1(\tau) - a_{20}z_2(\tau) \end{cases} \tag{21}$$

is exponentially convergent with respect to the origin $(0, 0)$. Where $a_{10}>0$ and $a_{20}>0$ are constants.

**Proof.** If the Lyapunov function candidate is selected as $V(z_1(\tau), z_2(\tau))=(a_{10}z_1^2(\tau)+z_2^2(\tau))/2$, then the result of the convergence can be obtained. $\square$

The process of convergent velocity for the system (21) is shown in Fig. 4.

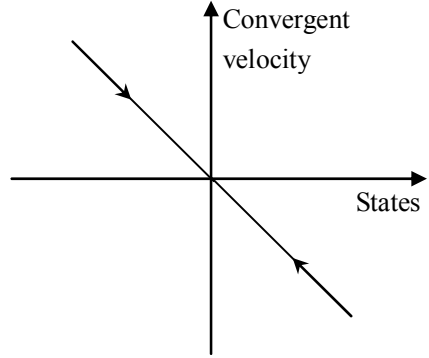

Fig.4. Convergent velocity in linear form

From Fig. 4, the convergent velocities of the states are rapid when they are far away from the equilibrium point. However, the convergent velocities are slow when they approach the equilibrium point. It is required to improve the process of convergent velocities when the states approach the equilibrium point.

**Lemma 2** (Convergence of nonlinear system)**.** System

$$\begin{aligned} \frac{d(z_1(\tau))}{d\tau} &= z_2(\tau) \\ \frac{d(z_2(\tau))}{d\tau} &= -a_{11}sig(z_1(\tau))^{\frac{\alpha}{2-\alpha}} - a_{21}sig(z_2(\tau))^{\alpha} \end{aligned} \tag{22}$$

is finite time convergent with respect to the origin $(0, 0)$, i.e.,

$$z_1(\tau) \equiv 0 \text{ and } z_2(\tau) \equiv 0 \text{ for } \tau \geq \tau_s \tag{23}$$

where $0 < \alpha < 1$, $a_{11}$ and $a_{21}$ are positive constants, $\tau_s$ is time constant.

**Proof.** Let the Lyapunov function candidate be:

$$V(z_1(\tau), z_2(\tau)) = \frac{a_{11}(2-\alpha)}{2}|z_1(\tau)|^{\frac{2}{2-\alpha}} + \frac{1}{2}z_2^2(\tau) \tag{24}$$

Therefore, we have

$$\begin{aligned} \frac{d(V(z_1(\tau), z_2(\tau)))}{d\tau} &= a_{11}sig(z_1(\tau))^{\frac{\alpha}{2-\alpha}}z_2(\tau) + z_2(\tau)\left(-a_{11}sig(z_1(\tau))^{\frac{\alpha}{2-\alpha}} - a_{21}sig(z_2(\tau))^{\alpha}\right) \\ &= -a_{21}|z_2(\tau)|^{1+\alpha} < 0 \end{aligned} \tag{25}$$

Note that the uniqueness of the solution of the system can be obtained from [31]. Then based on LaSalle invariant theory, the system (22) is asymptotically stable. Therefore, based on the negative homogeneity [30, Theorem 3.1], we have that

$$\varepsilon^{r_2} z_2(\tau) = \varepsilon^{r_1+k} z_2(\tau)$$

$$-a_{11} sig\left(\varepsilon^{r_1} z_1(\tau)\right)^{\frac{\alpha}{2-\alpha}} - a_{21} sig\left(\varepsilon^{r_2} z_2(\tau)\right)^{\alpha} = \varepsilon^{r_2+k}\left(-a_{11} sig\left(z_1(\tau)\right)^{\frac{\alpha}{2-\alpha}} - a_{21} sig\left(z_2(\tau)\right)^{\alpha}\right) \tag{26}$$

Therefore we have that

$$r_2 = r_1 + k, r_1 \frac{\alpha}{2-\alpha} = r_2 \alpha = r_2 + k \tag{27}$$

i.e.,

$$r_1 = k \frac{2-\alpha}{\alpha-1}, r_2 = k \frac{1}{\alpha-1} \tag{28}$$

Therefore, system (22) is finite time convergent with respect to the origin (0, 0), i.e.,

$$z_1(\tau) \equiv 0 \text{ and } z_2(\tau) \equiv 0 \text{ for } \tau \geq \tau_s \tag{29}$$

**Remark 3.** From [32], system

$$\frac{d(z_1(\tau))}{d\tau} = z_2(\tau)$$

$$\frac{d(z_2(\tau))}{d\tau} = -sig\left(z_1(\tau)\right)^{\alpha_1} - sig\left(z_2(\tau)\right)^{\alpha_2} \tag{30}$$

is finite time convergent with respect to the origin (0, 0), i.e.,

$$z_1(\tau) \equiv 0 \text{ and } z_2(\tau) \equiv 0 \text{ for } \tau \geq \tau_s \tag{31}$$

where $0 < \alpha_2 < 1$, and $1 > \alpha_1 > \alpha_2 / (2 - \alpha_2)$.

The process of convergent velocity for the system (22) or (30) is shown by solid line in Fig. 2.

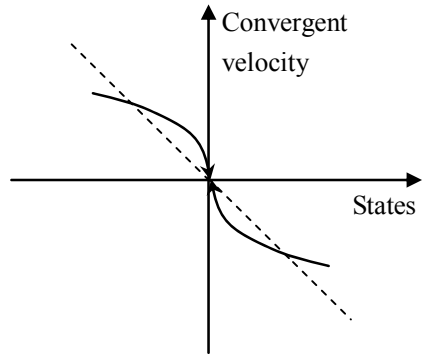

Fig.5. Convergent velocity in nonlinear form

From Fig. 5, although the convergent velocities of the states are rapid when they approach the equilibrium point, the convergent velocities of the states are slow when they are far away from the equilibrium point. Therefore, it is required to improve the convergent velocities whether the states are near to the equilibrium point or not, i.e., rapid convergence should be guaranteed at all times.

**Lemma 3** (Convergence of rapid-convergent system). System

$$\frac{d(z_1(\tau))}{d\tau} = z_2(\tau)$$

$$\frac{d(z_2(\tau))}{d\tau} = -a_{10} z_1(\tau) - a_{11} sig\left(z_1(\tau)\right)^{\frac{\alpha}{2-\alpha}} - a_{20} z_2(\tau) - a_{21} sig\left(z_2(\tau)\right)^{\alpha} \tag{32}$$

is finite time convergent with respect to the origin (0, 0), i.e.,

$$z_1(\tau) \equiv 0 \text{ and } z_2(\tau) \equiv 0 \quad \text{for} \ \tau \geq \tau_s \tag{33}$$

where $0 < \alpha < 1$, $a_{10}$, $a_{20}$, $a_{11}$ and $a_{21}$ are positive constants, $\tau_s$ is time constant.

**Proof.** We select the Lyapunov function candidate as:

$$V(z_1(\tau), z_2(\tau)) = \frac{a_{11}(2-\alpha)}{2}|z_1(\tau)|^{\frac{2}{2-\alpha}} + \frac{1}{2}\left(a_{10}z_1^2 + z_2^2(\tau)\right) \tag{34}$$

Therefore, we have

$$
\begin{aligned}
\frac{d\left(V(z_1(\tau), z_2(\tau))\right)}{d\tau} &= a_{11}sig\left(z_1(\tau)\right)^{\frac{\alpha}{2-\alpha}}z_2(\tau) + z_2(\tau)\left(-a_{10}z_1(\tau) - a_{11}sig\left(z_1(\tau)\right)^{\frac{\alpha}{2-\alpha}} - a_{20}z_2(\tau) - a_{21}sig\left(z_2(\tau)\right)^{\alpha}\right) \\
&\quad + a_{10}z_1(\tau)z_2(\tau) \\
&= -a_{21}|z_2(\tau)|^{1+\alpha} - a_{20}z_2^2(\tau) < 0
\end{aligned}
\tag{35}
$$

Note that the unique of the solution of the system can be obtained from [16]. Then based on LaSalle invariant theory, the system (32) is asymptotically stable.

From [30], $-a_{10}z_1(\tau) - a_{20}z_2(\tau)$ is locally uniformly bounded, whereas the right-hand side of the nominal model (22) is continuous and globally homogeneous of degree $q = k < 0$ with respect to dilation $r = (r_1, r_2)$. Hence, the condition $k + r_2 \leq 0$, required by Theorem 3.2 in [30], is satisfied, and Theorem 3.2 in [30] is applicable to the globally equiuniformly asymptotically stable (32). By applying Theorem 3.2 in [30], (32) is thus globally equiuniformly finite time stable. The proof of Lemma is completed. □

**Remark 4.** From [32] and [30], system

$$
\begin{aligned}
\frac{d(z_1(\tau))}{d\tau} &= z_2(\tau) \\
\frac{d(z_2(\tau))}{d\tau} &= -a_{10}z_1(\tau) - a_{11} - sig\left(z_1(\tau)\right)^{\alpha_1} - a_{20}z_2(\tau) - a_{21}ssig\left(z_2(\tau)\right)^{\alpha_2}
\end{aligned}
\tag{36}
$$

is finite time convergent with respect to the origin (0, 0), i.e.,

$$z_1(\tau) \equiv 0 \text{ and } z_2(\tau) \equiv 0 \quad \text{for} \ \tau \geq \tau_s \tag{37}$$

where $0 < \alpha_2 < 1$, and $1 > \alpha_1 > \alpha_2/(2-\alpha_2)$, $a_{10}$, $a_{20}$, $a_{11}$ and $a_{21}$ are positive constants, $\tau_s$ is time constant..

The process of convergent velocity for the system (32) or (36) is shown by solid line in Fig. 6. From Fig. 6, we can find that the convergent velocities keep rapid whether the states approach to the equilibrium point or not, that is to say, the rapid-convergent velocities can be guaranteed at all times.

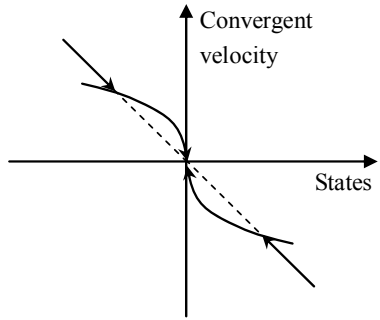

Fig.6. Convergent velocity in rapid-convergent form

**5.2 Proof of Theorem 2.**

Taking Laplace transformation of linear differentiator (17), we have

$$sX_1(s) = X_2(s)$$
$$\varepsilon^2 sX_2(s) = -a_{10}(X_1(s) - V(s)) - a_{20}\varepsilon X_2(s) \quad (38)$$

Therefore,

$$\varepsilon^2 sX_2(s) = -a_{10}\left(\frac{X_2(s)}{s} - V(s)\right) - a_{20}\varepsilon X_2(s) \quad (39)$$

Then, we have

$$\frac{sV(s)}{X_2(s)} = 1 + \frac{\varepsilon^2 s^2 + a_{20}\varepsilon s}{a_{10}} \quad (40)$$

i.e.,

$$\frac{X_2(s)}{V(s)} = \frac{s}{1 + \dfrac{\varepsilon^2 s^2 + a_{20}\varepsilon s}{a_{10}}} \quad (41)$$

It is clear that

$$\lim_{\varepsilon \to 0} \frac{X_2(s)}{V(s)} = \lim_{\varepsilon \to 0} \frac{s}{1 + \dfrac{\varepsilon^2 s^2 + a_{20}\varepsilon s}{a_{10}}} = s \quad (42)$$

It means that $x_2(t)$ approximates the derivative $\dot{v}(t)$. This ends the proof of Theorem 2. $\square$

**Remark 4.** In fact, the differentiator (17) is equivalent to the one presented in [4]. Select

$$x_1 = w_1 - \varepsilon a_2 w_2 / a_1, x_2 = w_2 \quad (43)$$

we have

$$\dot{w}_1 = w_2 - a_2(w_1 - v(t))/\varepsilon$$
$$\dot{w}_2 = -a_1(w_1 - v(t))/\varepsilon^2 \quad (44)$$
$$y = w_2$$

From (41), we find that the differentiation is achieved over a limited frequency range $1/\varepsilon$. State $x_2(t)$ of differentiator (17) is the output of a concatenated ideal 1-order differentiator and a low-pass filter of order 2. By increasing the differentiation order 2, noise is more attenuated, but bandwidth becomes smaller than the usual one. The controllable canonical form (17) seems to be so interesting when the differentiator is used in closed-loop configurations. In addition, we see that $v(t)$ just appears in differentiator (17), so a great amount of eventual additive noise shall be eliminated because of the presence of the successive 2 integrators.

Linear differentiator (17) is the boundary layer of system (21), and the convergent velocities of the state variables are slow in the nonlinear region of the system dynamics, the lagging phenomenon is inevitable. This is confirmed by Fig. 4.

For linear differentiator (17), we also find that when the tracking error is large, i.e., far from the origin, its behavior is strong. On the other hand, when tracking error is small, i.e., around the origin, its behavior is weak.

In order to remove the slow convergence in the nonlinear region of the system dynamics, we present an algorithm of nonlinear differentiator (18), and we given the convergence proof of differentiator (18) in the following.

### 5.3 Proof of Theorem 3.

From Lemma 2, we know that system (22) is finite-time-convergent with respect to the origin with a finite time $T_f$.

And from Theorem 1 in [34], we have the result (20) for nonlinear differentiator (18).

**Remark 5.** Based on finite-time convergence of (22) (in Remark 3), we have another simple differentiator as follow:

$$\dot{x}_1 = x_2$$
$$\varepsilon^2 \dot{x}_2 = -sig(x_1 - v(t))^{\alpha_1} - a_{21}sig(\varepsilon x_2)^{\alpha_2} \qquad (38)$$
$$y = x_2$$

where $0 < \alpha_2 < 1$, and $1 > \alpha_1 > \alpha_2/(2 - \alpha_2)$.

Differentiators (36) and (38) are the boundary layers of system (18) and (30), respectively. However, rapid convergence is local from Fig. 2. We can find that when the tracking error of (18) is large, i.e., far from the origin, the gains in (18) are small. On the other hand, when the tracking error is small, i.e., around the origin, the gains are large. Therefore, the behavior of nonlinear differentiator (18) is weak far from the origin, and contrarily, the behavior around the origin is strong (See Fig. 5).

The differentiator (15) can not only reduce sufficiently chattering phenomenon of derivative estimation, but also its dynamical performances are improved by adding linear correction terms to the nonlinear ones. In the following, we will give the convergence of differentiator (15).

### 5.4 Proof of Theorem 1.

From Lemma 3, we know that system (32) is finite-time-convergent with respect to the origin with a finite time $T_f$.

And from Theorem 1 in [34], we have the result (16) for rapid-convergent differentiator (15).

**Remark 6.** Based on finite-time convergence of (36) (in Remark 4), we have another simple differentiator as follow:

$$\dot{x}_1 = x_2$$
$$\varepsilon^2 \dot{x}_2 = -a_{10}(x_1 - v(t)) - a_{11}sig(x_1 - v(t))^{\alpha_1} - a_{20}\varepsilon x_2 - a_{21}sig(\varepsilon x_2)^{\alpha_2} \qquad (39)$$
$$y = x_2$$

where $0 < \alpha_2 < 1$, and $1 > \alpha_1 > \alpha_2/(2 - \alpha_2)$, $a_{10}$, $a_{20}$, $a_{11}$ and $a_{21}$ are positive constants.

Differentiators (15) and (39) are the boundary layers of system (32) and (36), respectively. We can find that differentiator (15) has the rapid convergence at all times from Fig. 6, and due to the simple construct, the computation time is very short. Moreover, because of the continuity of the system, no chattering phenomena happen. This type of differentiator is adapted to the systems requiring rapid convergence.

### 6. Robustness analysis of finite-time-convergent differentiator

Here, we will give the robustness of finite-time-convergent differentiator (11).

**Theorem 4:** For finite-time-convergent differentiator (11), if there exists a noise in signal $v(t)$, that is

$v(t) = v_0(t) + d(t)$, where $v_0(t)$ is the desired second-order derivative signal, and $d(t)$ is a bounded noise and is

satisfied with $|d(t)| \le h_d$, we have

$$x_i - v^{(i-1)}(t) = O\left(\varepsilon^{\rho_\gamma - i + 1}\right)$$

under the condition that $\varepsilon < \left(\dfrac{c}{2M\delta}\right)^{\frac{1}{\rho}}$ and $h_d < \left(\dfrac{c-2\varepsilon^\rho M\delta}{2M\bar{a}}\right)^{\frac{1}{\rho_d}}$.

**Proof:** Differentiator (11) can be written as

$$\frac{dx_1}{dt} = x_2$$
$$\cdots$$
$$\frac{dx_{n-1}}{dt} = x_n \tag{40}$$
$$\varepsilon^n \frac{dx_n}{dt} = f\left(x_1 - v(t) - d(t), \varepsilon x_2, \cdots, \varepsilon^{n-1} x_n\right)$$

Let

$$e_1 = x_1 - v_0(t), e_2 = x_2 - v_0^{'}(t), \cdots, e_n = x_n - v_0^{(n)}(t)$$

The error system is

$$\frac{de_1}{dt} = e_2$$
$$\cdots$$
$$\frac{de_{n-1}}{dt} = e_n \tag{41}$$
$$\varepsilon^n \frac{de_n}{dt} = f\left(e_1 - d(t), \varepsilon e_2 + \varepsilon\frac{dv}{dt}, \ldots, \varepsilon^{n-1}e_2 + \varepsilon^{n-1}\frac{d^{n-1}v}{dt^{n-1}}\right) - \varepsilon^n\frac{d^n v}{dt^n}$$

System (41) can be written as

$$\frac{de_1}{dt/\varepsilon} = \varepsilon e_2$$
$$\cdots$$
$$\frac{d\varepsilon^{n-2}e_{n-1}}{dt/\varepsilon} = \varepsilon^{n-1}e_n \tag{42}$$
$$\frac{d\varepsilon^{n-1}e_n}{dt/\varepsilon} = f\left(e_1 - d(t), \varepsilon e_2 + \varepsilon\frac{dv}{dt}, \ldots, \varepsilon^{n-1}e_n + \varepsilon^{n-1}\frac{d^{n-1}v}{dt^{n-1}}\right) - \varepsilon^n\frac{d^n v}{dt^n}$$

Let

$$\tau = \frac{t}{\varepsilon},$$
$$z_1(\tau) = e_1(t), z_2(\tau) = \varepsilon e_2(t), \cdots, z_n(\tau) = \varepsilon^{n-1}e_n(t), \tag{43}$$
$$z = [z_1 \quad \cdots \quad z_n]^{\mathrm{T}}, \bar{d}(\tau) = d(t)$$

we have $z = \Xi(\varepsilon)e$, where $\Xi(\varepsilon) = diag\{1, \varepsilon, \cdots, \varepsilon^{n-1}\}$. From (43), system (42) can be transferred as

$$\frac{dz_1}{d\tau} = z_2$$
$$\dots$$
$$\frac{dz_{n-1}}{d\tau} = z_n \tag{44}$$
$$\frac{dz_n}{d\tau} = f\left(z_1 - \bar{d}(\tau), z_2 + \frac{dv}{d\tau}, \dots, z_n + \frac{d^{n-1}v}{d\tau^{n-1}}\right) - \frac{d^n v}{d\tau^n}$$

We find that system (44) is the perturbation system of (10). Select Lyapunov be $(V \circ z)(\tau)$ and let $z^0 = \begin{bmatrix} z_2 & \cdots & z_n \end{bmatrix}$.

From Assumption 2, we know that $(V \circ z)(\tau)$ is Lipschitz, and its' Lipschitz constant is $M$. Therefore, we have

$$D^+(V \circ z)(\tau) = \frac{\partial V}{\partial z}(z)\left[z^0 \quad f\left(z_1, z_2 + \frac{dv}{d\tau}, \dots, z_n + \frac{d^{n-1}v}{d\tau^{n-1}}\right) - \frac{d^n v}{d\tau^n}\right]^{\mathrm{T}}$$
$$= \frac{\partial V}{\partial z}(z)\left[z^0 \quad f(z_1, z_2, \dots, z_n)\right]^{\mathrm{T}}$$
$$+ \frac{\partial V}{\partial z}(z)\left\{\left[z^0 \quad f\left(z_1 - \bar{d}(\tau), z_2 + \frac{dv}{d\tau}, \dots, z_n + \frac{d^{n-1}v}{d\tau^{n-1}}\right) - \frac{d^n v}{d\tau^n}\right]^{\mathrm{T}} - \left[z^0 \quad f(z_1, z_2, \dots, z_n)\right]^{\mathrm{T}}\right\} \tag{45}$$
$$\leq \dot{V} + M\bar{a}\left(\left|\bar{d}(\tau)\right|^{\rho_d} + \sum_{i=1}^{n-1}\left|\frac{d^i v}{d\tau^i}\right|^{\rho_i}\right) + M\left|\frac{d^n v}{d\tau^n}\right|$$
$$= \dot{V} + M\left[\bar{a}\left|\bar{d}(\tau)\right|^{\rho_d} + \bar{a}\sum_{i=1}^{n-1}\varepsilon^{i\rho_i}\left|\frac{d^i v}{dt^i}\right|^{\rho_i} + \varepsilon^n\left|\frac{d^n v}{dt^n}\right|\right]$$
$$\leq \dot{V} + M\left[\bar{a}h_d^{\rho_d} + \bar{a}\sum_{i=1}^{n-1}h_i^{\rho_i}\varepsilon^{i\rho_i} + h_n\varepsilon^n\right]$$

where $h_i > 0$ is the up boundness of $\left|dv_0^i/dt^i\right|$. Set $\rho = \min\{i\rho_i\}$. From Assumption 3, we know $\rho_d \in (0, 1]$ and $\delta = \bar{a}\sum_{i=1}^{n-1}h_i^{\rho_i} + h_n$. Therefore, we can get

$$D^+(V \circ z)(\tau) \leq \dot{V} + \varepsilon^\rho M\left[\bar{a}\sum_{i=1}^{n-1}h_i^{\rho_i} + h_n\varepsilon^n\right] \tag{46}$$
$$= \dot{V} + \varepsilon^\rho M\delta + M\bar{a}h_d^{\rho_d}$$

It is required that the perturbation in system (44) be continuous, and this requirement hold in each $[t_{j-1}, t_j), j = 1, \dots, k+1$. Therefore, there exists a constant $\Gamma > 0 (\Gamma > T_f)$, for $t_j/\varepsilon \geq \tau \geq t_{j-1}/\varepsilon + \Gamma\left(z(t_{j-1}/\varepsilon)\right)$, $j = 1, \dots, k+1$, satisfied with [30, Theorem 5.2 and Proposition 2.4]

$$\|z(\tau)\| \leq \frac{(V(z(\tau)))^{1-\theta}}{rc(1-\theta)} \leq \frac{\left(\frac{2\varepsilon^\rho M\delta + 2M\bar{a}h_d^{\rho_d}}{c}\right)^{\frac{1-\theta}{\theta}}}{rc(1-\theta)} \tag{47}$$

$\theta$ can be chosen to be sufficiently small and $\frac{1-\theta}{\theta}$ can be sufficiently large, therefore, we have

$$\frac{2\varepsilon^\rho M\delta + 2M\bar{a}h_d^{\rho_d}}{c} < 1 \tag{48}$$

Therefore, $z(\tau)$ is sufficiently small. From (48), we have

$$h_d < \left( \frac{c - 2\varepsilon^\rho M \delta}{2M\overline{a}} \right)^{\frac{1}{\rho_d}}$$

(49)

Therefore, when $\varepsilon < \left( \frac{c}{2M\delta} \right)^{\frac{1}{\rho}}$ and $h_d < \left( \frac{c - 2\varepsilon^\rho M \delta}{2M\overline{a}} \right)^{\frac{1}{\rho_d}}$, noise can be restrained sufficiently.

## 7. Reduction of peaking phenomenon

In rapid-convergent differentiator, peaking phenomena happen due to the infinity $\varepsilon$. In order to reduce peaking phenomena sufficiently, we choose $\varepsilon$ as

$$1/\varepsilon = \begin{cases} \mu t & if \quad 0 \le t \le t_{\max} \\ \mu t_{\max} & \text{otherwise} \end{cases}$$

(50)

where $\mu$ and $t_{max}$ are chosen according to the desired maximum error that depends on the value of $\gamma_{max} = \mu \, t_{\max}$.

## 8. Simulations and experiments
### 8.1 convergences of three types of systems
Because (17), (18) and (15) are the boundary layer of systems (21), (22) and (32), respectively, we give the convergences of (21), (22) and (32) firstly, and their convergences are shown in Fig. 7.

$$\begin{cases} \dot{z}_1 = z_2 \\ \dot{z}_2 = -z_1 - z_2 \end{cases}, \quad \begin{cases} \dot{z}_1 = z_2 \\ \dot{z}_2 = -sig(z_1)^{0.5} - sig(z_2)^{0.5} \end{cases}, \quad \begin{cases} \dot{z}_1 = z_2 \\ \dot{z}_2 = -z_1 - sig(z_1)^{0.5} - z_2 - sig(z_2)^{0.5} \end{cases}$$

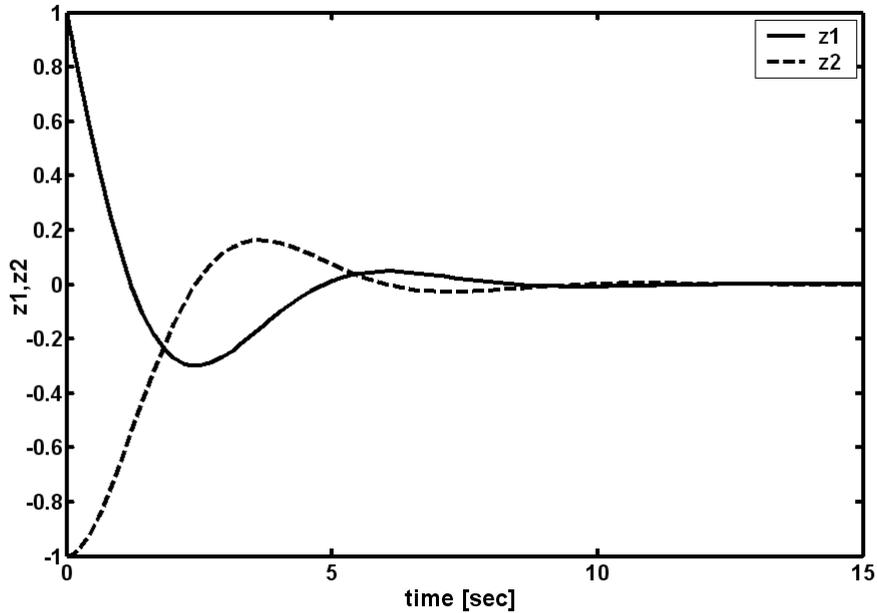

(7-a) linear convergence

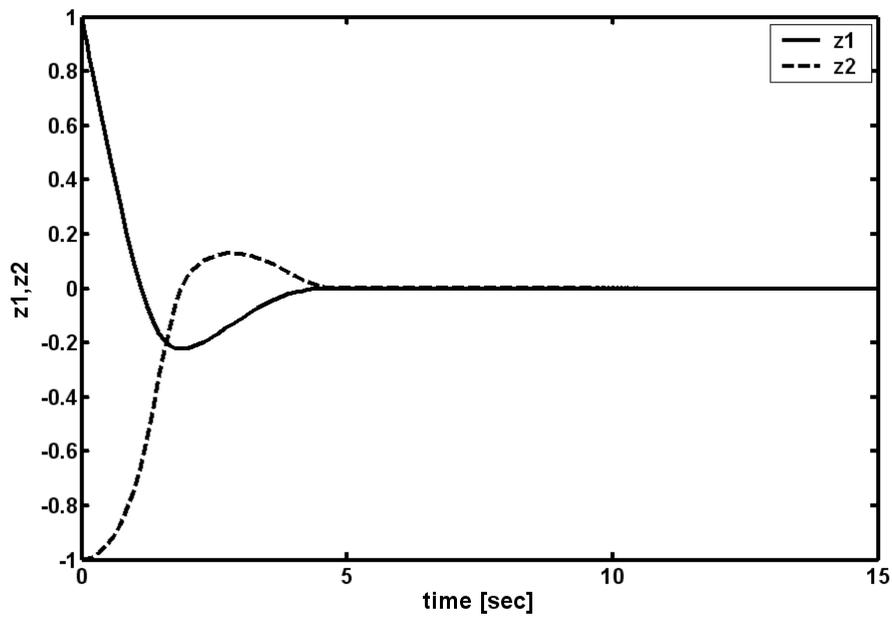

(7-b) nonlinear convergence

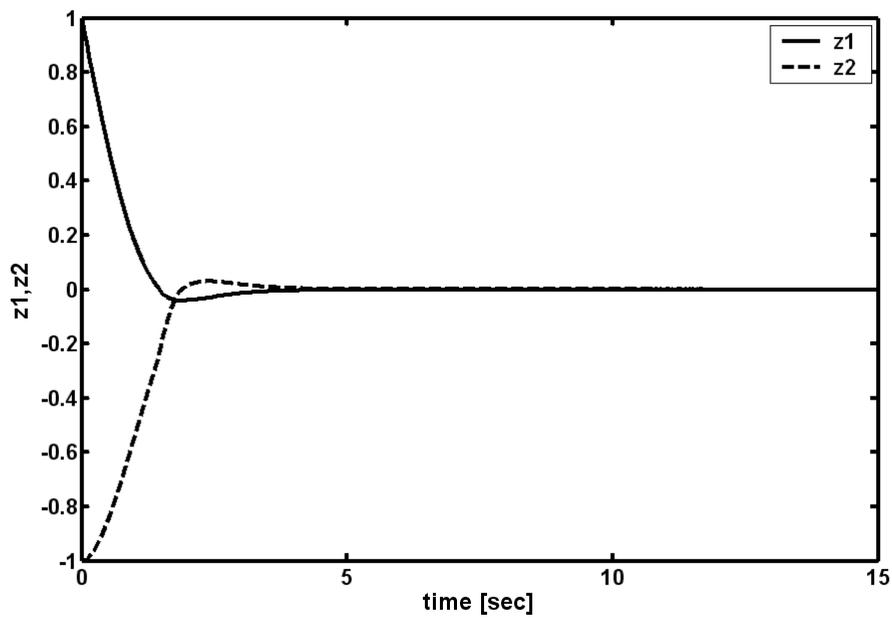

(7-c) linear-nonlinear convergence

Fig. 7. The convergences of three systems

## 8.2 Three types of differentiator

In the following, we select the functions of $\sin t$ and triangular wave, respectively, as the input signal $v(t)$.

1) Linear differentiator (differentiator (17)). Parameters: $a_{10}$=5, $a_{11}$=2, $\varepsilon$=1/300. The derivative estimations of linear differentiator are shown in Fig. 8.

$$\dot{x}_1 = x_2$$
$$\frac{1}{300^2}\dot{x}_2 = -5(x_1 - v(t)) - \frac{2}{300}x_2$$
$$y = x_2$$

2) Nonlinear differentiator (differentiator (18)). Parameters: $\varepsilon = 1/300$, $a_{11} = 5$, $a_{21} = 2$, $\alpha = 0.5$. The derivative estimations of nonlinear differentiator are shown in Fig. 9.

$$\dot{x}_1 = x_2$$
$$\frac{1}{300^2}\dot{x}_2 = -5 sig(x_1 - v(t))^{0.5} - 2 sig\left(\frac{x_2}{300}\right)^{0.5}$$
$$y = x_2$$

3) Rapid-convergent differentiator (differentiator (15)). Parameters: $a_{10} = 5$, $a_{11} = 2$, $a_{20} = 0.5$, $a_{21} = 0.5$, $\varepsilon = 1/300$, $\alpha_1 = \alpha_2 = 0.5$.

$$\dot{x}_1 = x_2$$
$$\frac{1}{300^2}\dot{x}_2 = -5(x_1 - v(t)) - 0.5 sig(x_1 - v(t))^{0.5} - 2\frac{x_2}{300} - 0.5 sig\left(\frac{x_2}{300}\right)^{0.5}$$
$$y = x_2$$

The derivative estimations of rapid-convergent differentiator are shown in Fig. 10.

From the simulations above, for the representative input signal $v(t) = \sin t$, we can see that the convergent time $t_s = 0.025$s in linear differentiator based on singular perturbation technique, computation time: 0.8s, however, the vibrating phenomenon is obvious; the convergent time $t_s = 0.18$s in nonlinear differentiator based on singular perturbation technique, computation time: 1s, the lagging phenomenon happen; the convergent time $t_s = 0.01$s in rapid-convergent differentiator based on singular perturbation technique, computation time: 1.2s, the effect of tracking has high precision and no chattering phenomenon happen.

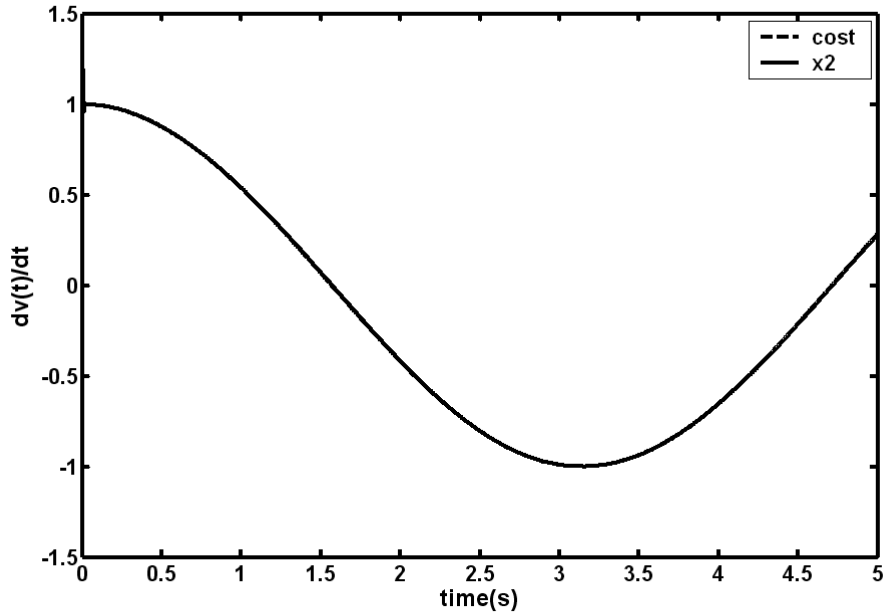

(8-a) Tracking "sin$t$" wave

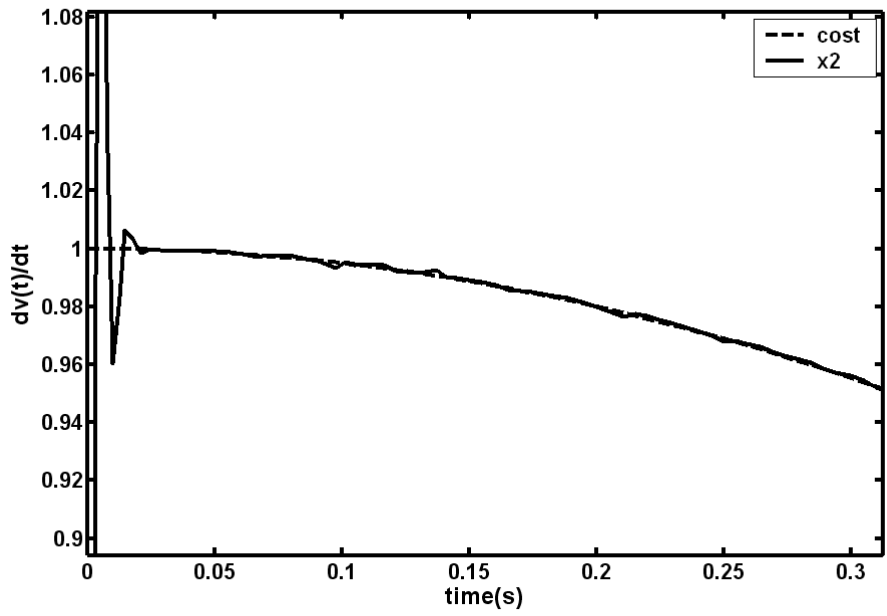

(8-b) The magnified figure of (8-a)

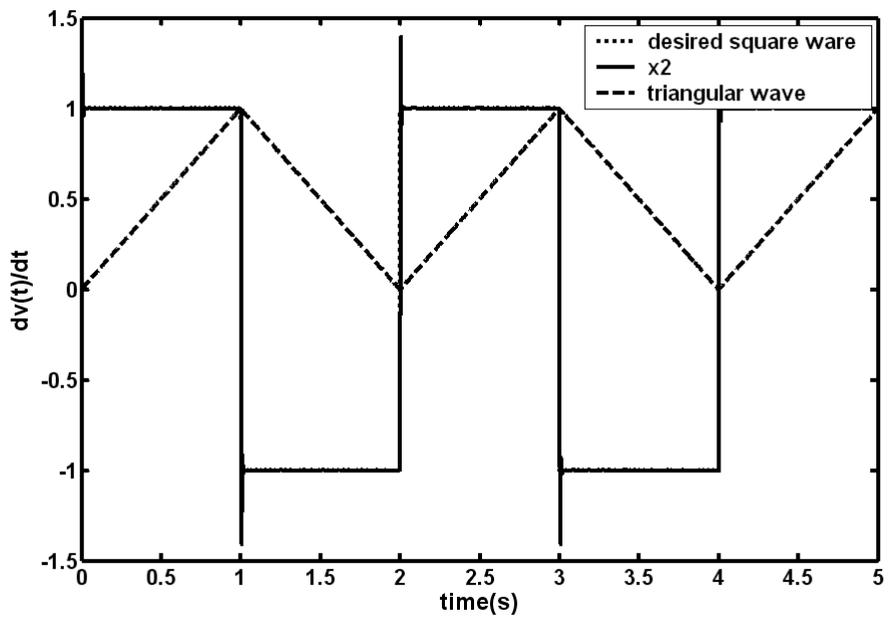

(8-c) Tracking "Triangular" wave

Fig.8. Linear differentiator

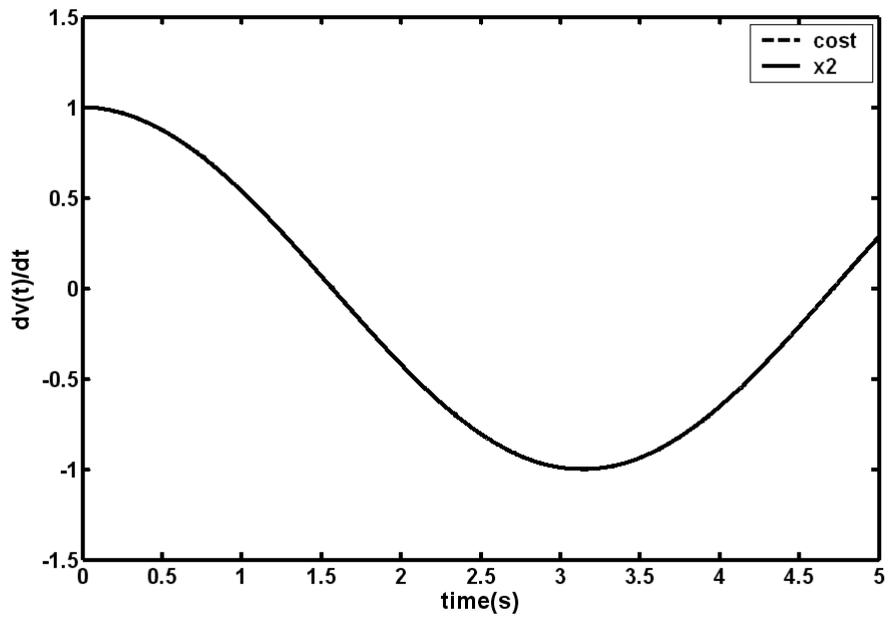

(9-a) Tracking "sinr" wave

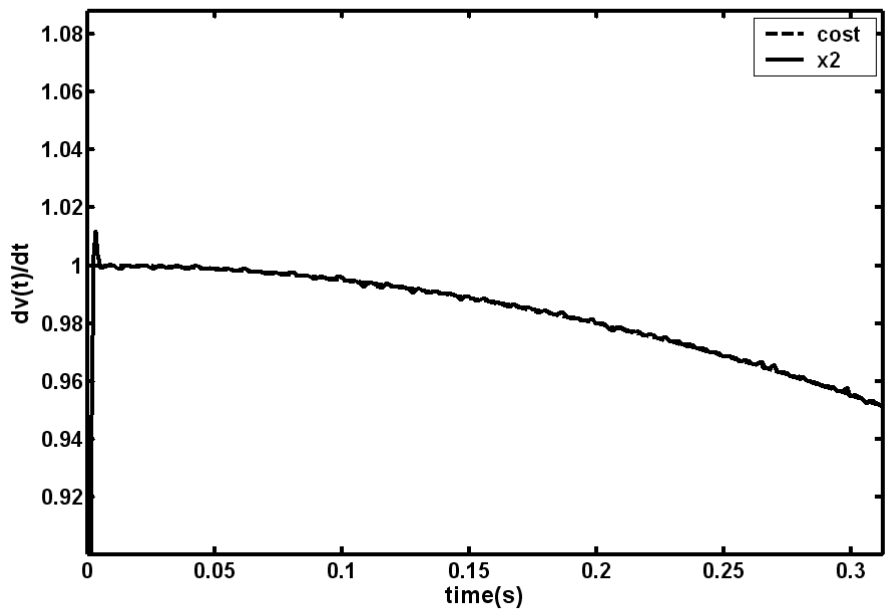

(9-b) The magnified figure of (9-a)

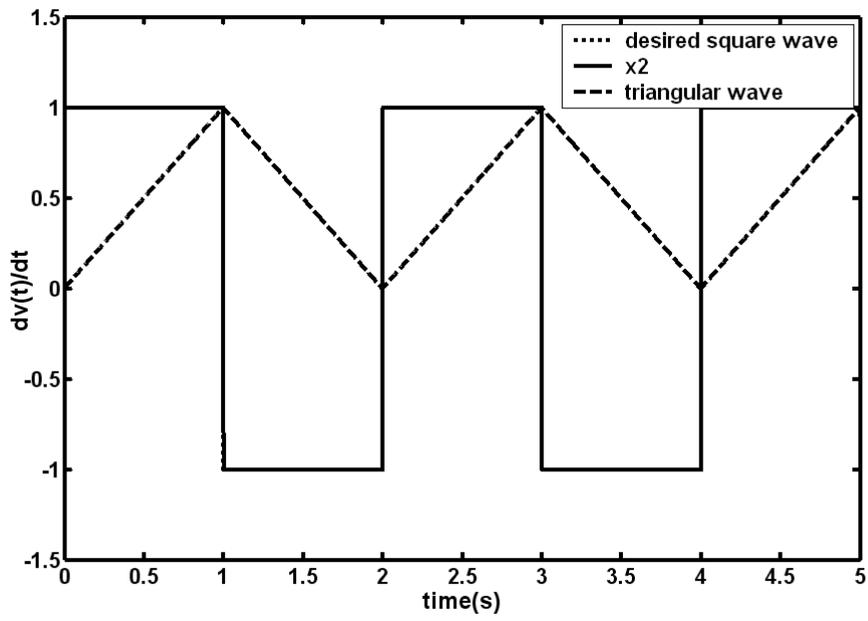

(9-c) Tracking "Triangular" wav

Fig.9. Nonlinear differentiator

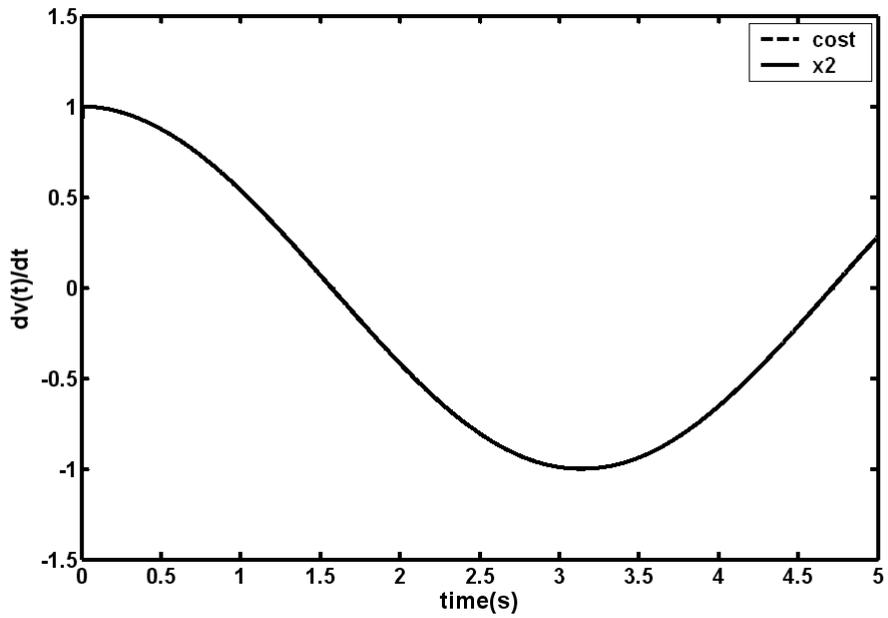

(10-a) Tracking "sinr" wave

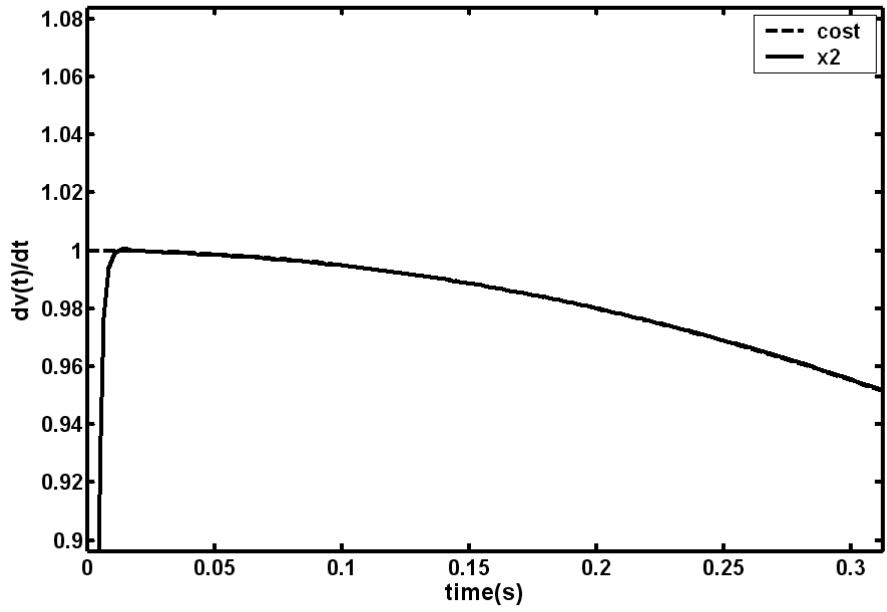

(10-b) The magnified figure of (10-a)

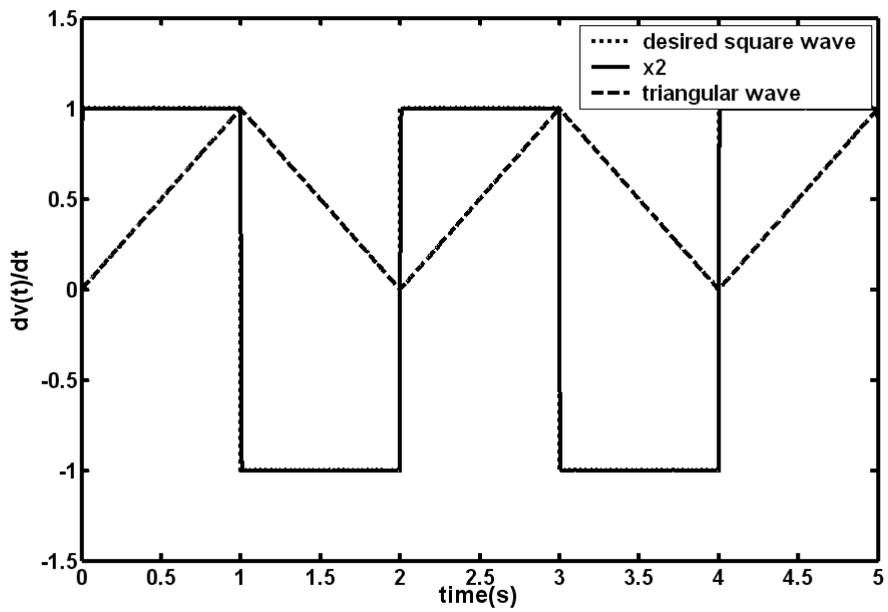

(10-c) Tracking "Triangular" wave

Fig.10. Rapid-convergent differentiator

## 83. Estimation of uncertainty with noise

If there exists noise in an uncertain system, we can use the differentiator to estimate the uncertainty, and noise can be restrained sufficiently.

We consider the following uncertain system

$$\dot{x} = -x + u + \delta$$
$$y_{un} = x + \xi$$

where $x \in \mathbf{R}$ is the state, $u$ is the control input, $y_{un}$ is the measurement output, $\delta$ is the uncertainty, $\xi$ is the noise. We use the differentiator to estimate the uncertainty $\delta$, i.e.,

$$\delta \approx \dot{\hat{x}} + x - u$$

Suppose $u$=0.1sin$t$, and $\delta = \cos t$. We use the following rapid-convergent differentiator:

$$\dot{x}_1 = x_2$$
$$\dot{x}_2 = -0.05 \times 45^2 \left( x_1 - y_{un} \right) - 0.015 \times 45^2 \, sig \left( x_1 - y_{un} \right)^{0.6}$$
$$\quad - 0.3 \times 45 x_2 - 0.015 \times 45^2 \, sig \left( \frac{x_2}{45} \right)^{0.6}$$
$$y = x_2$$

The measurement output with noise is shown in Fig. 11, and the estimation of uncertainty by rapid-convergent differentiator is shown in Fig. 12.

We can find that the uncertainty can be estimated with satisfying precision in spite of noise existing in the measurement output.

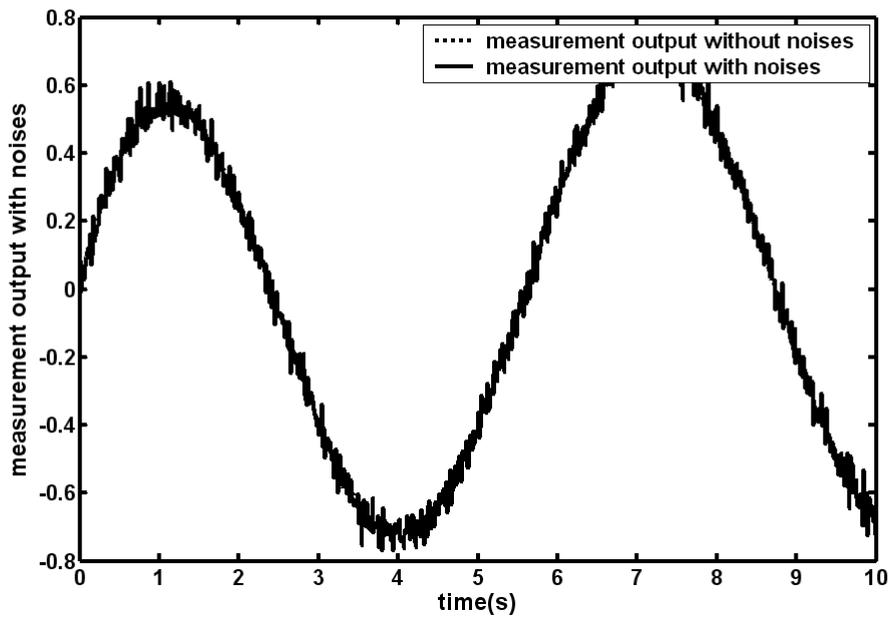

Fig. 11. Measurement output with noise

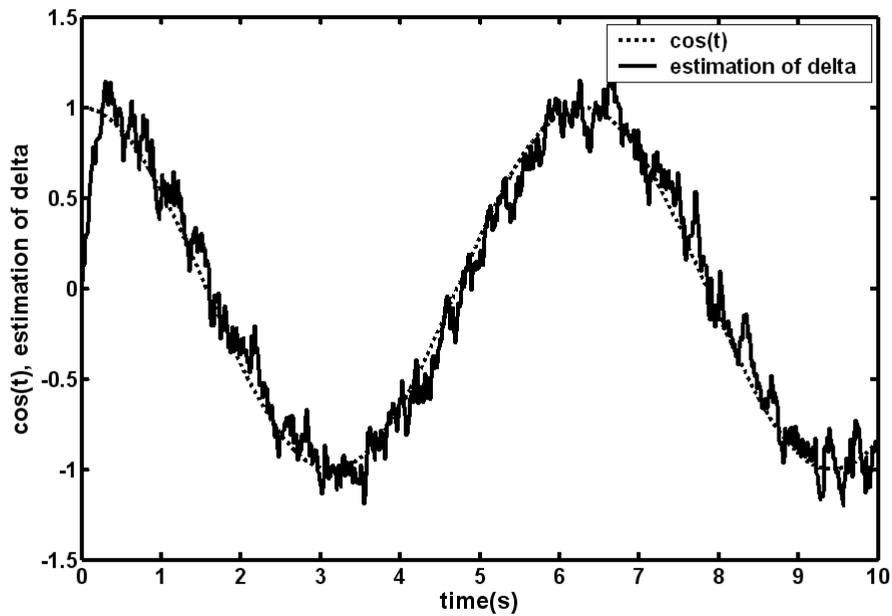

Fig. 12. The estimation of uncertainty

**8.4 Experiment for rapid-convergent differentiator**

In the following, we use rapid-differentiator to estimate, respectively, the rotor acceleration from rotor speed and velocity of convertor voltage from convertor voltage for a servo system with direct current double loop speed driving. The simulink of control system is given in Fig. 13, and the equipments of servo system are shown in Fig. 14.

The rotor speed and convertor voltage are set sinusoidal and "triangular" waveforms, respectively, and the rapid-convergent differentiator is compiled in DSP6713 (Digital Signal Processor 6713), the differential equation is carried out by the method of 4-order RungeKutta. For the apparatus in Fig. 14, through an oscillograph, we can obtain the derivative signals shown in Fig. 15.

From Fig. 15, we can find that though the input signals have much noise due to the effect of mechanical vibration and other reasons, the derivative estimations obtained from rapid-convergent differentiator still have satisfying quality, and the noises are restrained sufficiently by the differentiator. Not only chattering phenomenon and noises can be reduced sufficiently, but also dynamical performances are improved effectively.

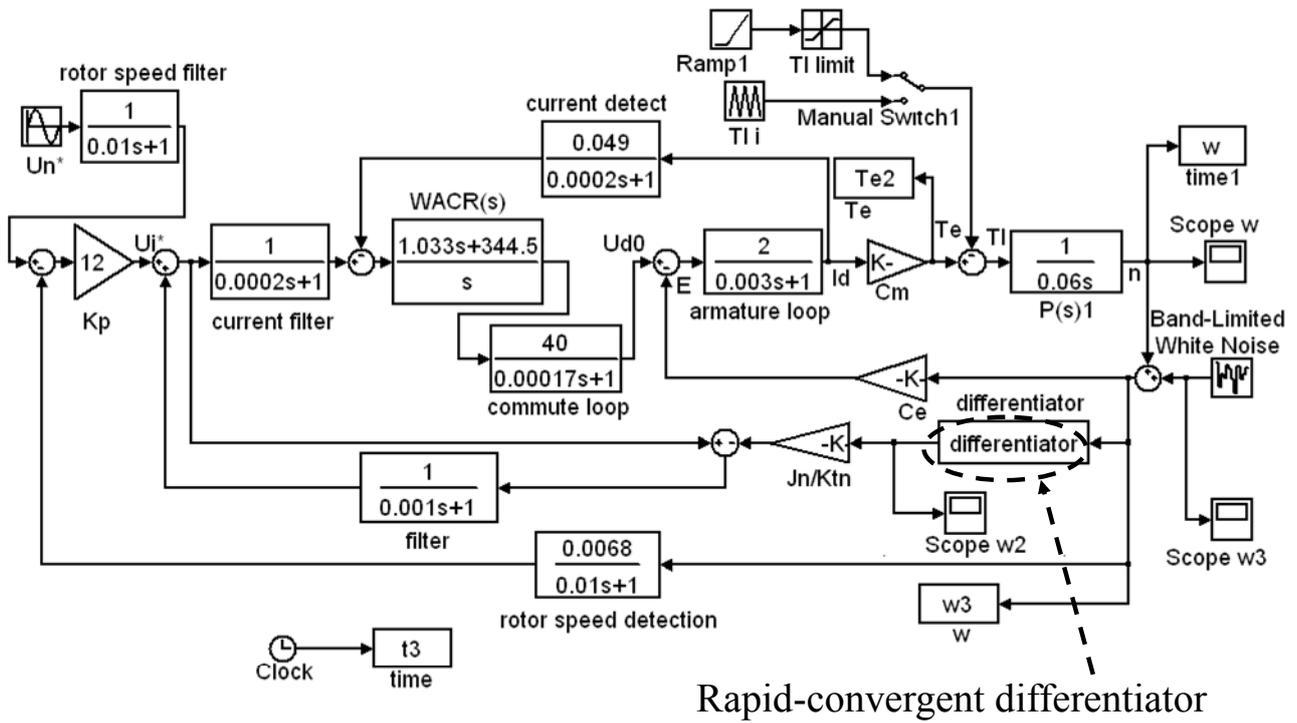

Rapid-convergent differentiator

Fig. 13. Simulink of servo system

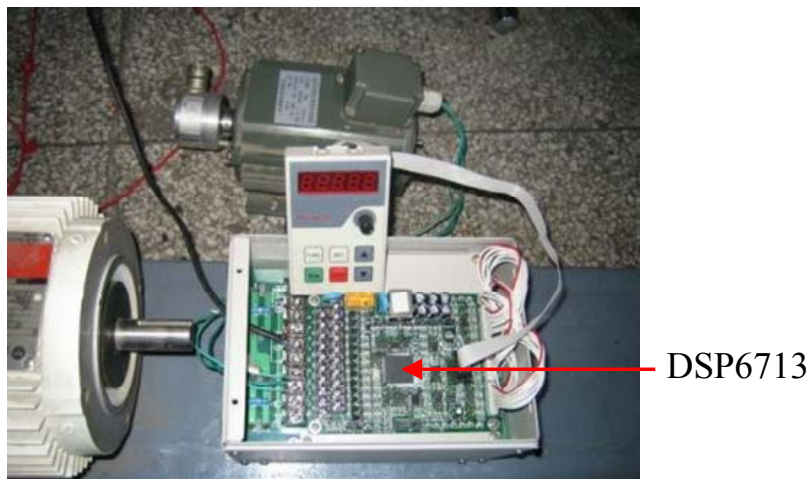

DSP6713

Fig. 14. The equipment of servo system

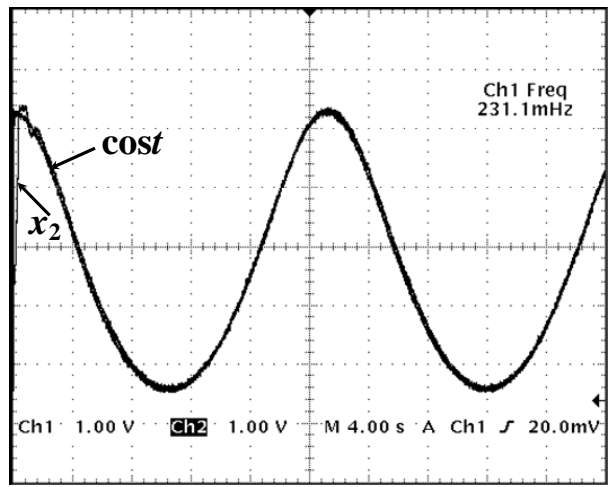
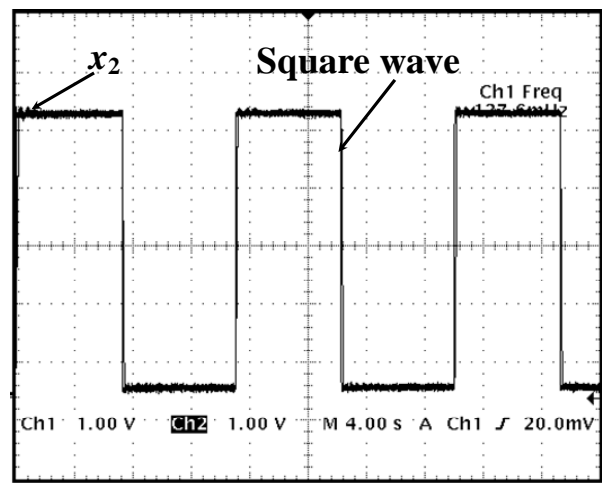

<div align="center">(15-a ) Tracking "sin<i>t</i>" wave          (15-b) Tracking "Triangular" wave</div>

<div align="center">Fig. 15. An experiment for rapid-convergent differentiator</div>

## 9. Conclusions

In this paper, we present an algorithm of rapid-convergent differentiator based on singular perturbation technique and compare it with other approaches of differentiation. Because of its continuous structure, and consists of linear and nonlinear parts, not only chattering phenomenon and noises can be reduced sufficiently, but also dynamical performances are improved effectively. However, the parameters in rapid-convergent differentiator are not optimal, and the differentiator is not suitable to delayed signal. Our future work is to design a robust differentiator for delayed signal and use adaptive methods to optimize differentiators.